Title

# Secondary cratering from Rheasilvia as the possible origin of Vesta's equatorial troughs

Authors
**Naoyuki Hirata** [a], *

* Corresponding Author E-mail address: hirata@tiger.kobe-u.ac.jp

**Authors' affiliation**
[a] Graduate School of Science, Kobe University, Kobe, Japan.

**Proposed Running Head: Origin of Vesta's equatorial troughs**
Editorial Correspondence to:
Dr. Naoyuki Hirata
Kobe University, Rokkodai 1-1 657-0013
Tel/Fax +81-7-8803-6566

Abstract

Asteroid 4 Vesta has a set of parallel troughs aligned with its equator. Although previous evaluations suggest that it is of shock fracturing tectonic origin, we propose that the equatorial troughs can be created by secondary cratering from the largest impact basin, Rheasilvia. We calculated the trajectories of ejecta particles from Rheasilvia by considering Vesta's rapid rotation. As a result, we found that secondary craters should be parallel to the latitude. In particular, if we assume that ejecta particles are launched at an initial launch velocity of approximately 350-380 m/s and a launch angle of 25 °, the parallel equatorial troughs, the Divalia Fossae, can be suitably explained by secondary cratering. This model works well on objects, such as Haumea, Salacia, and Chariklo, but not on Mercury, the Moon, and regular satellites.

Plain language abstract

Asteroid 4 Vesta has a set of parallel troughs aligned with its equator. We propose that the equatorial troughs can be created by secondary cratering from Rheasilvia, the largest impact basin of Vesta. We calculated the trajectories of ejecta particles from Rheasilvia by considering Vesta's rapid rotation. As a result, we found that secondary craters should be parallel to the latitude. In particular, the pattern of troughs indicates that ejecta particles were launched at an initial launch velocity of approximately 350-380 m/s and a launch angle of 25 °.

Key points

- We propose a new mechanism for the formation of equatorial troughs on Vesta
- We calculated the distribution of ejecta particles launched from Rheasilvia
- We found that secondary cratering from Rheasilvia matches with the equatorial troughs

## 1. Introduction

Asteroid 4 Vesta is one of the largest objects in the asteroid belt and is believed to be a remnant intact protoplanet from the earliest epoch of solar system formation and parent body of howardite-eucrite-diogenite (HED) meteorites [e.g., Russell and Raymond 2011]. NASA's Dawn spacecraft closely approached Vesta between July 2011 and August 2012, sending numerous high-resolution images to the Earth and revealing the nature of this asteroid. For example, the spatially resolved mineralogy of the surface reflects the composition of the HED meteorites, confirming the formation of Vesta's crust via the melting of a chondritic parent body. Further, Vesta's mass, volume, and gravitational field indicate that Vesta is a differentiated object with sufficient internal melting to segregate iron [Russell et al. 2012]. Vesta has two large impact basins, Rheasilvia and Veneneia, on its south pole. The Rheasilvia basin, with a diameter of ~500 km, is the largest impact crater on Vesta, while the Veneneia basin, with a diameter of ~400 km, is the second largest [Schenk et al. 2012]. The Veneneia basin is older and more degraded than Rheasilvia, and half of Veneneia's basin floor overlaps with Rheasilvia [Schenk et al. 2012]. The formation of the Rheasilvia is estimated to be approximately 1 Ga based on an asteroid flux-derived chronology [Schenk et al. 2012; Marchi et al. 2012; O'Brien et al. 2014; Yingst et al. 2014] or 3.5 Ga based on a lunar-derived chronology [Schmedemann et al. 2014]. On the other hand, the preserved parts of the Veneneia basin are heavily damaged by the Rheasilvia impact, making it difficult to estimate its formation age [Schenk et al. 2012].

One of the notable features of Vesta is the large-scale trough systems occurring in the equatorial region (Divalia Fossae) and northern hemisphere (Saturnalia Fossae) of Vesta (Fig. 1a) [Jaumann et al. 2012]. The most notable set of troughs on Vesta occurs along the equator (mainly between 20°S and 0°S), encircles roughly 2/3 of Vesta's equatorial region [Jaumann et al. 2012], and does not cut the Vestalia Terra [Buczkowski et al. 2012, 2014]. These troughs, collectively called Divalia Fossae, are wide, flat-floored, and bounded by steep scarps, including muted troughs, grooves, and pit crater chains [Buczkowski et al. 2012; Jaumann et al. 2012]. The largest of these troughs is Divalia Fossa, which is 465 km long, 5 km deep, and 14.5-22 km wide [Buczkowski et al. 2012; Schäfer et al. 2014]. The other set of troughs on Vesta,

Saturnalia Fossae, is found in the northern hemisphere. These troughs extend to the northwest, and are oriented ~30° from the equator. The largest of these troughs is Saturnalia Fossa [Scully et al. 2014]. The poles of Divalia Fossae agreed with the basin center of Rheasilvia, while those of Saturnalia Fossae agreed with the basin center of Veneneia, suggesting that the formation of the troughs is related to the formation of the two impact basins [Jaumann et al. 2012; Buczkowski et al. 2012]. In fact, the crater age of Divalia Fossae is consistent with that of the Rheasilvia basin, indicating that they can be considered coeval within the bounds of the uncertainties [Cheng et al. 2021]. In addition, the walls of Saturnalia Fossae display gentler slopes and rounded edges, which implies considerable infilling and heavy cratering, thereby suggesting it was older than Divalia Fossae [Jaumann et al. 2012; Buczkowski et al. 2012], which is consistent with the stratigraphic relationship between Rheasilvia and Veneneia.

These troughs have been interpreted to be graben systems, induced by normal faulting at the moment of impact events [Buczkowski et al. 2012; Jaumann et al. 2012]; in other words, the troughs are of shock fracture origin. This view was tested using numerical modeling and experiments. For example, Ivanov and Melosh [2013] estimated the shock wave decay, escaped material volume, and depth of excavation of Rheasilvia using a two-dimensional hydrocode and reproducing the Rheasilvia crater size and shape. Bowling et al. [2014] calculated the strains induced by the impact stress wave at the Rheasilvia impact event to gain insight into the locations of strains and the expected mode of deformation. Stickle et al. [2015] demonstrated that, based on experimental and numerical results, the offset angle (between the orientation of the troughs and the Rheasilvia basin center) is a natural consequence of oblique impacts on a spherical target. Similar to Vesta, various objects in the solar system have trough systems that may be related to impact-induced shock fractures or normal faulting. The Argyre basin on Mars has concentric graben and ridge systems around it [e.g. Hiesinger and Head 2002]. Multi-ring trough systems on Ganymede and Callisto are created by normal faulting induced by impacts [e.g., McKinnon and Melosh 1980; Melosh 1982; Hirata et al. 2020]; grooves on Phobos are created by shock fractures induced by the Stickney impact event [e.g., Fujiwara and Asada 1983; Asphaug and Melosh 1993]; and grooves identified on asteroids, Gaspra,

Ida, Eros, and saturnian satellites might have been created by impacts [Veverka et al. 1994; Sullivan et al. 1996; Prockter et al. 2002; Morrison et al. 2009].

Instead of the shock fracture origin, we propose that secondary cratering launched at the Rheasilvia event could create large-scale troughs. The trajectories of crater ejecta launched from rapidly rotating objects are considerably affected by rotation. For example, Schmedemann et al. (2018) examined secondary impact distributions from Rheasilvia taking the Coriolis effect into an account, and compared a color map of Vesta and the distribution. The landing locations of ejecta from the Stickney crater on Phobos are expected to be elongated toward the west owing to satellite rotation resulting from the Coriolis effect [e.g., Dobrovolskis and Burns 1980; Davis et al. 1981; Thomas 1998]. In particular, linear grooves and crater chains on Phobos can be explained by the reaccretion of ejecta from impacts on Phobos [Nayak and Asphaug 2016]. Hirata and Ikeya [2021] and Ikeya and Hirata [2021] demonstrated that ejecta launched from a rapidly rotating asteroid tend to accumulate along the equator of the asteroid owing to the Coriolis effect, which could explain the bluer spectral units along the equatorial ridge of Ryugu. In this work, we examined the trajectories of ejecta launched from Rheasilvia by considering Vesta's rapid rotation.

Radial troughs, such as secondary crater chains or radial sculptures, appear around large impact basins, such as the Imbrium and Orientale basins on the Moon and the Caloris basin on Mercury [e.g., Spudis 1993]. Over the past few decades, numerous studies have been conducted on radial structures around impact basins. The Fra Mauro Formation, recognized during pioneering telescopic studies of the Moon [Gilbert 1893] and known as the Apollo 14 landing site, is composed of parallel troughs with a width of approximately 12 km and is considered a prominent example of a radial sculpture centered in the Imbrium basin [e.g., Head and Hawke 1975]. Around the Orientale Basin, there are numerous crater chains or clusters radial to the basin center, extending a radial distance of $6R$ ($R$ = 465 km) from the basin rim crest [e.g., Guo et al. 2018]. The Van Eyck Formation, characterized by a series of grooves, ridges, and crater chains with a width of 5–30 km, extends radially from the Caloris at least hemispherical scale on Mercury [McCauley et al. 1981; Fassett et al. 2009]. Most of these radial structures are proposed to be caused by secondary cratering due to low-angle ejection from the crater [e.g., Baldwin 1963; Wilhelms 1976; Head 1976; Fassett et al. 2009; Guo et al. 2018]. We expect that in Vesta, secondary cratering from Rheasilvia should occur along the equator owing to the fast rotation period of Vesta, and the so-called radial sculpture should appear as equatorial parallel troughs.

## 2. Method

We calculated the landing locations of the ejecta particles launched from the Rheasilvia. To compute the ejecta trajectories, we followed the method described by Hirata et al. [2021], as described below. Although it is not clear whether these launch velocity distributions are appropriate in this type of global-scale impact basin, we consider them meaningful as first-order estimates.

The initial launch velocity of a particle in a gravity-dominated regime is given by [Housen and Holsapple 2011]

$$v_{ej} = C_1 \left( H_1 \sqrt[3]{\frac{4\pi}{3}} \right)^{-\frac{2+\mu}{2\mu}} \sqrt{gR_c} \left( \frac{x}{R_c} \right)^{-\frac{1}{\mu}} \left( 1 - \frac{x}{n_2 R_c} \right)^p \quad , (n_1 a \leq x \leq n_2 R_c) \,, (1)$$

where $v_{ej}$ is the initial velocity of the launched particle, $R_c$ is the apparent crater radius, $x$ is the distance between the ejection position and crater center, $g$ is the surface gravity, $a$ is the projectile radius (here, we assume $a = 0$), and the rest are scaling constants that depend on target materials (Table 1). In addition, we assumed $\theta$= 20°, 25°, 30°, and 45° as the ejecta launch angle (Fig. 1b). Here, we assume that the launch angle is constant with respect to x. The case in which it is not a constant is shown in Appendix A. Housen and Holsapple [2011] provided eight sets (C1–C8) of scaling constants, and we utilized C1 (target is water), C4 (dry sand), and C6 (glass micro-spheres) (Table 1). The three velocity distribution models did not differ markedly from each other (Fig. 2a). Notably, C2, C3, C7, and C8 are in the strength regime, and C5 is identical to C4. We assumed the crater center to be at 75°S 301°E, and $x$ was defined by the point at which the ejecta particle crossed the original surface (Fig. 1b). Distances, $x$ and $R_c$, are defined as the great circle distances on the object. Notably, $R_c$ is the apparent crater radius and $n_2 R_c$ is the crater rim radius. Because the Rheasilvia basin has a diameter of ~500 km [Schenk et al. 2012], the following is employed: $n_2 R_c$ =250 km.

The trajectory of an ejecta particle is solved using the equation of motion in a rotating reference frame [Scheeres et al. 2002]:

$$\ddot{\boldsymbol{r}} + 2\boldsymbol{\Omega} \times \dot{\boldsymbol{r}} + \boldsymbol{\Omega} \times \boldsymbol{\Omega} \times \boldsymbol{r} = -\frac{GM}{|r|^3}\boldsymbol{r} \quad , (2)$$

where $\boldsymbol{r}$ is the position vector of the ejecta particle relative to an asteroid-centered body-fixed frame and $\boldsymbol{\Omega}$ is the rotation vector. The z-axis was set as the rotational axis of the object. We assumed a Vesta-equivalent spherical object, with a radius of $R$= 525.4 km, a mass of $M = 2.59 \times 10^{20}$ kg, and a rotation period of $T$=5.342 hours [Russell et al. 2012]. Given the initial launch position and velocity of the particle, given by Eq. (1), the trajectory of the particle is obtained by numerically integrating Eq. (2). The initial launch velocity in Eq. (1) is defined as the velocity relative to the surface of Vesta. At first glance, the equation does not include the velocity component of the rotation of Vesta. However, because the calculations were performed in the rotating reference frame, the ejecta particles had an initial velocity, including the component of the rotation velocity of Vesta in the inertial reference frame. We therefore do not need to add the velocity component of the rotation of Vesta into Eq. (1). We determined that ejecta particle that reached below the asteroid surface collided with the asteroid, and ejecta particle that reached an altitude greater than the Hill radius (i.e. the radius of the Hill sphere [Hamilton and Burns 1991]) of the asteroid escaped from the asteroid.

## 3. Result

In general, particles launched at low speeds have a short travelling time (the time from launch to landing), fall in the vicinity of the crater, and are barely affected by the Coriolis force, whereas those launched at sufficiently high speeds have a long travelling time and are sufficiently affected by the Coriolis force. As an example, Fig. 2c and 2d show the relationship between the longitude and latitude of the landing locations and the initial launch velocity of the ejecta particles indicated by the light blue line shown in Fig. 3c, 4c, and 5c. A particle with a launch velocity of less than 200 m/s is barely affected by the Coriolis force (i.e., the longitude of the landing locations is barely affected), whereas one with a launch velocity higher than 270 m/s is affected and folded toward the west (Fig. 2c). A particle launched at 200 m/s has a travelling time of approximately 1200 s, whereas that launched at 270 m/s has a travelling time of 3800 s (Fig. 2b). In general, the escape velocity of Vesta is 360 m/s, however, in the case of Fig. 2cd, a particle launched at < 380 m/s does not escape. This is because the launch direction affects the actual escape velocity of the particles owing to the velocity component of asteroid rotation.

Figs. 3, 4, and 5 show the landing locations of the particles launched from Rheasilvia for C1, C4, and C6, respectively. A line with the same color indicates the landing locations of the particles ejected in the same direction. The trails of the landing locations of particles sufficiently far away from the basin were aligned with the latitudes. For example, landing locations of ejecta particles focus on 45°N–90°N when the initial launch angle is $\theta$=45°, 10°S–30°N when $\theta$=30°, 30S°-10N° when $\theta$=25°, and 50°S–10°S when $\theta$=20°. Shallower launch angles tend to focus on the southern latitudes. This result does not seem to depend on launch velocity models, C1, C4, and C6. Because a set of equatorial troughs is observed between the equator and 20°S, an initial launch angle of $\theta$=25° can most suitably explain the distribution of the observed troughs. In this case, ejecta particles landing around the equator (within 30S°-10N°) have an initial launch velocity between approximately 350 m/s and 380 m/s (Fig. 2d). The particles launched at 350 m/s had a travelling time of 30000 s, which is equivalent to 1.5 Vesta days. On the other hand, if we assume $T$=∞ (non-rotation), focusing toward the opposite point of the crater center can be seen; however, the horizontal lines shown in Figs. 3, 4, and 5 do not occur (Fig. 6).

## 4. Discussion

### 4.1 Case of Vesta

The results show that secondary impactors with an initial launch velocity of approximately 350-380 m/s and a launch angle of 25° would accumulate along the equator. It is not unusual for the launch angle to be slightly lower than that in the general case (45°) because most of the radial-sculptures/troughs/secondary-craters on the Moon and Mercury are considered to be caused by secondary cratering due to low-angle ejection [e.g., Baldwin 1963; Wilhelms 1976; Head 1976; Fassett et al. 2009; Guo et al. 2018]. The maximum width of the Divalia Fossa is 22 km, which roughly corresponds to 4% of the Rheasilvia basin diameter. On the Moon, Mercury, Dione, and Rhea, the maximum size of the secondary craters is approximately 4% [Melosh 2011; Schenk et al. 2020], therefore, the width of the troughs is consistent with the secondary crater origin.

The volume of ejecta launched faster than $v_{ej}$ is given by [Housen and Holsapple 2011]

$$V = kC_1^{3\mu} \left(H_1 \sqrt[3]{\frac{4\pi}{3}}\right)^{-\frac{6+3\mu}{2}} R_c^3 \left(\frac{v_{ej}}{\sqrt{gR_c}}\right)^{-3\mu}, \quad (3)$$

where $k$ is a constant (see Table 1). Assuming that ejecta particles with an initial launch velocity between 350 and 380 m/s create equatorial troughs on Vesta, the volumes of ejecta that create the equatorial troughs are $6.6 \times 10^4$ km³ (Target C1), $6.6 \times 10^4$ km³ (Target C4), and $7.8 \times 10^4$ km³ (Target C6). The scaling links the size of a crater to the size of the projectile [Holsapple and Housen 2007; Melosh 2011]; assuming an impact velocity of 300 m/s, the radius of a crater is approximately three times as large as that of the projectile, and therefore, the ratio of the volume of the crater to the projectile would be approximately 27. Therefore, the total volume of secondary craters along the equator would be $1.8 \times 10^6$ km³. The depression of Divalia Fossa, 18 km wide on average, 465 km long, and 5 km deep, has up to a volume of $4.2 \times 10^4$ km³. In summary, the volume of ejecta particles that accumulate along the equator can create 43 Divalia-scale troughs.

Figs. 2-5 show trails of ejecta particles encircling Vesta's equator; however, this model does not suggest that Divalia-style troughs completely cover the equator. A possible explanation for the fact that troughs on Vesta are less pronounced around the Vestalia Terra may be that the initial launch velocity or launch azimuth of the ejecta particles are not homogenous (Appendix B). Otherwise, it is possible that the rigidity of the Vestalia Terra may explain the paucity of Divalia-style troughs in terms of a tectonic origin. The Vestalia Terra is thought to be comprised of stronger material than the rest of Vesta's equatorial region and associated with a mascon created by the head of a magmatic plume (Raymond et al. 2013; Raymond et al. 2017).

There are a few possible explanations for the origin of Saturnalia Fossae. (1) Our calculations (Figs. 3-5) show that ejecta particles launched at low speed from trails extending to the northwest, oriented several tens of degrees from the equator. These ejecta particles were launched at between 270 and 300 m/s and were moderately affected by the Coriolis force. Some of these diagonal trails may match Saturnalia Fossae. In this case, the secondary impactors should be derived from either Rheasilvia or Veneneia. (2) If the north or south pole of Vesta is located at the center of the Veneneia basin when Veneneia was created, Saturnalia Fossae should be aligned with the latitude line. Similar to Divalia Fossae, it is possible that Saturnalia fossae were created by secondary cratering aligning along the latitude lines launched from the Veneneia. In this case, ejecta forming Saturnalia Fossae should be launched at $\theta=30°$, because (i) the colatitude (i.e. angle formed by the two points relative to the center of Vesta) between the Rheasilvia basin center and Divalia Fossae was 90° on Vesta, while that between Veneneia basin center and Saturnalia Fossae is 110 ° and (ii) as shown in Figs. 3-5, the latitude that the landing locations of ejecta particles focus on becomes more distant from the basin center than do the ejecta particles with a steeper launch angle. In addition, this case requires a scenario in which the poles of Vesta migrated after Veneneia impact.

Finally, we briefly discuss the geomorphological properties of the troughs in Vesta. Fig. 7a-d show Dawn FC images of the equatorial troughs on Vesta. Previous studies have examined their geomorphological properties and interpreted them as tectonic fractures, such as grabens. For example, Buczkowski et al. [2012] and Jaumann et al. [2012] reported that the troughs are wide, flat-floored, and bounded by steep scarps, including muted troughs, grooves, and pit crater chains. Schäfer et al. [2014] and Yingst et al. [2014] reported that they are primarily composed of parallel running troughs and ridges (Fig. 7cd). Cheng and Klimczak [2022] reported that most of the troughs have bowl-shaped cross-sectional geometries and a distinct flat floor is not the most common trough geomorphology and they are heavily degraded, and there is no diagnostic evidence for fault traces of graben or joint walls, although they concluded that the troughs are of tectonic origin. On the other hand, Buczkowski et al. [2012] showed several examples of fault linkages and fault slip movement in Divalia and Saturnalia Fossa. Although these interpretations are reasonable, the possibility of a secondary crater origin has not been evaluated in previous studies. Radial sculptures around the Caloris basin (Fig. 7e) are composed of parallel linear ridges and depressions, which are considered a continuous linear trail of impacts created by a stream of ejecta particles, therefore, they are very dense and overlap each other and often do not resemble the morphology of a single chain crater. Most sculptures had bowl-shaped cross-sectional geometries and were bounded by ridges. These features are similar to those of the equatorial troughs of Vesta. It has also been reported that pit crater chains align with the equator (curly brackets in Fig. 7ab) [Buczkowski et al. 2012; Jaumann et al. 2012]. Two of the pit chains are named: Robigalia Catena (Fig. 7a) and Albalonga Catena (Fig. 7b). One of the Divalia Fossae troughs shows a direct transition into the Albalonga Catena pit crater chain; the trough narrows toward its tips and directly transitions into a pit crater chain, which aligns with smaller pits beyond the end of the trough [Buczkowski et al. 2014; Cheng and Klimczak 2022]. These pit chains resemble the nature of secondary craters from Orientale (Fig. 7f), which exhibit a set of parallel chains of bowl-shaped craters linearly-connected to each other. In addition, it is known that there are many clusters of small craters on Vesta based on spatial randomness analyses [Cheng et al. 2021]; in fact, the equatorial region of Vesta appears as many clusters of small craters (white arrows in Fig. 7a-d). Because clusters of secondary craters are created away from their primary crater, it is natural that there are many clusters around the equator if secondary impactors from Rheasilvia accumulated on the equator. However, it should be noted that there is significant existing literature suggesting that pit crater chains are tectonic features, not secondary crater chains [e.g. Wyrick et al. 2004; Ferrill et al. 2004, 2011; Smart et al. 2011; Martin et al. 2017; Frumkin and Naor 2019; Whitten and Martin 2019; Wyrick and Buczkowski 2022]. For example, pit crater chains on Mars are observed to be spatially correlated with normal faults and have been identified as transitioning into graben [Wyrick et al. 2004], comparable to what is observed in Albalonga Catena.

### 4.2 Comparisons with other objects

The question arises as to why only Vesta has a large-scale trough system along its equator. We consider this topic based on a dimensionless parameter: the ratio of $V_{trough}$ (the volume of ejecta that create troughs along latitude) to $V_{object}$ (a volume of the object). The volume of the ejecta launched faster than a given velocity, as described by Eq. (3) and Table 1. From Eq. (3), the volume of ejecta launched between $v_{min}$ and $v_{max}$ can be described by

$$V = kC_1{}^{3\mu} \left(H_1 \sqrt[3]{\frac{4\pi}{3}}\right)^{-\frac{6+3\mu}{2}} R_c{}^3 \left[\left(\frac{v_{min}}{\sqrt{gR_c}}\right)^{-3\mu} - \left(\frac{v_{max}}{\sqrt{gR_c}}\right)^{-3\mu}\right] \quad . (4)$$

By setting the minimum and maximum values to $v_{min}$ and $v_{max}$ of the initial launch velocity of an ejecta particle that creates a trough along its latitude line, we can obtain $V_{trough}$. The ratio of $V_{trough}$ to the volume of the object $V_{object} = 4\pi R^3/3$ is written as

$$V_{trough}/V_{object} = kC_1{}^{3\mu} H_1{}^{-\frac{6+3\mu}{2}} \left(\frac{4\pi}{3}\right)^{-2-\frac{\mu}{2}} \left(\frac{R_c}{R}\right)^3 \left[\left(\frac{\sqrt{gR_c}}{v_{min}}\right)^{3\mu} - \left(\frac{\sqrt{gR_c}}{v_{max}}\right)^{3\mu}\right] . (5)$$

Eq. (5) is a dimensionless quantity; thus, we can compare cases for various objects. Here, we must define $v_{min}$ and $v_{max}$ uniformly. We assume $v_{max}$ to be the escape velocity of the object, although the actual escape velocity of the particles is affected by the velocity component of object rotation. The escape velocity means the minimum launch velocity of an ejecta particle that reaches an altitude of the Hill radius of the object. However, $v_{min}$ is ambiguous and difficult to determine uniquely. From the results of Sections 2 and 3, we empirically assumed $v_{min}$ to be the launch velocity of the ejecta particle whose travelling time is equal to one or two rotation periods of the object. By solving a simple equation of motion for a free-fall body launched from an object, these initial launch velocities were obtained for various asteroids or dwarf planets (Table 2). Table 2 lists the values of $v_{min}$, $v_{max}$, and $V_{trough}/V_{object}$ for various impact craters in the solar system. The values for one rotation period tended to be approximately twice as high as those for the two rotation periods. Of these impact craters, the value of $V_{trough}/V_{object}$ for Rheasilvia was the highest; therefore, Vesta has the greatest advantage in the creation of troughs along its latitude lines.

Although the rotation period of Ceres is short (9.07 hours), $V_{trough}/V_{object}$ for Kerwan, the largest crater on Ceres, is approximately one-hundredth of that of Rheasilvia. This might be because Kerwan is not a large crater relative to Ceres, only a quarter of the diameter of Ceres. In fact, Ceres does not exhibit large-scale trough systems. On the other hand, Ceres has a set of linear structures called Junina Catenae that have been proposed to be a secondary crater chain from the Urvara crater (Fig. 8a) [Schmedemann et al. 2017; Scully et al. 2017]. Junina Catenae are also aligned with the latitudes. We calculated the landing locations of ejecta particles from Urvara, Kerwan, and Yalode (Fig. 8b-d), which showed that the ejecta particles could accumulate along the Junina Catenae. Our model indicates that the Junina Catenae can be explained by any large impact crater, not only Urvara, but also Kerwan, Yalode, or any large craters (Fig. 8b-d). The Samhain Catenae, another set of linear structures on Ceres (Fig. 8a), does not match ejecta from Urvara and Kerwan, but they can match ejecta from Yalode. Therefore they may have had a secondary impact origin, although Scully et al. [2017] proposed that they may be the surface representation of buried faults, formed due to past interior activity.

Similarly, we calculated the landing locations of the ejecta particles from Caloris, Chicxulub, Orientale, and Hellas (Fig. 9). In the case of impact craters on Mars and the Earth, some of the ejecta particles landed along the latitude lines (Fig. 9bd); however, the value of $V_{trough}/V_{object}$ was small (Table 2) because the object rotation period was slightly long and $v_{min}$ was close to $v_{max}$. Although the Hellas basin is large, the $V_{trough}/V_{object}$ of Hellas is one hundredth that of Rheasilvia. In the case of Orientale or Caloris, the value of $V_{trough}/V_{object}$ becomes zero because their Hill radius is very small, their escape velocity is too small, their rotation periods are too long, and $v_{min}$ cannot be defined. In fact, the landing locations of ejecta particles from the Orientale or Caloris do not align with the latitude (Fig. 9ac).

In the case of the Moon, Mercury, and saturnian satellites, the value of $V_{trough}/V_{object}$ becomes zero (Table 2). Even if this type of object has structures along the latitude, our model does not explain these structures. For example, ray-like bright structures along latitudes have been found in Dione [Hirata and Miyamoto 2016]. The majority of bright structures on Dione are aligned radially from the Creusa crater; therefore, it is regarded as the ray system of the Creusa crater; however, the rest (ones along the latitude) are not aligned radially (Fig. 10a). At first glance, we consider that ray-like bright structures along the latitude of Dione may also be explained by rays of Creusa folded by the Coriolis force (Fig. 10b,c); however, this is impossible because the hill sphere of Dione is too small. Using Hill's equation instead of Eq. (2), the landing locations of ejecta particles do not align with the latitude (Fig. 10d).

In addition, for each object, we assume that there is an impact crater with a radius equal to the radius of the object ($R_c = R$), half of the object's radius ($R_c = 0.5R$), and a quarter of the radius of the object ($R_c = 0.25R$) and compared $V_{trough}/V_{object}$ for each object (Table 3 and Fig. 11). We then include objects that have never been observed by spacecraft. Distant objects have a slight advantage in forming troughs along their latitudes owing to their large Hill radii. In particular, Centaur or Trans-Neptunian objects, such as Haumea, Salacia, and Chariklo, have a greater advantage than Vesta (Fig. 11). If these objects have a large impact basin, we can predict that a radial sculpture around the large impact basin will become a sculpture along the latitude.

## 5. Conclusion

We calculated the trajectories of ejecta particles from Rheasilvia, assuming Housen and Holsapple's equation as an initial launch velocity distribution and considering asteroid rotation. Although our model does not rule out previous models, we demonstrate that the parallel equatorial troughs, Divalia Fossae, can be explained by secondary cratering, given a suitable initial condition. For example, when the initial launch velocity is approximately 350-380 m/s and the launch angle is 25 °, the trails of the landing locations of ejecta particles are deflected in the direction of rotation and parallelly aligned with latitudes between 30S°-10N°; this is consistent with the distribution of the equatorial troughs. In addition, the volume of the ejecta can create 43 Divalia-scale troughs. We predict that distant objects, such as Haumea, Salacia, and Chariklo, would have similar parallel troughs along their latitudes if the objects have large impact craters that are large relative to their diameter.

Acknowledgments

The author wishes to thank Leonard D. Vance and an anonymous reviewer for their helpful comments, which significantly improved the manuscript. This work was partly supported by the JSPS Grants-in-Aid for Scientific Research (Nos. 20K14538 and 20H04614) and the Hyogo Science and Technology Association.

Open Research

Data Availability Statement

The software for calculating ejecta trajectories from Rheasilvia is available at Zenodo (https://doi.org/10.5281/zenodo.7700522).

Figures

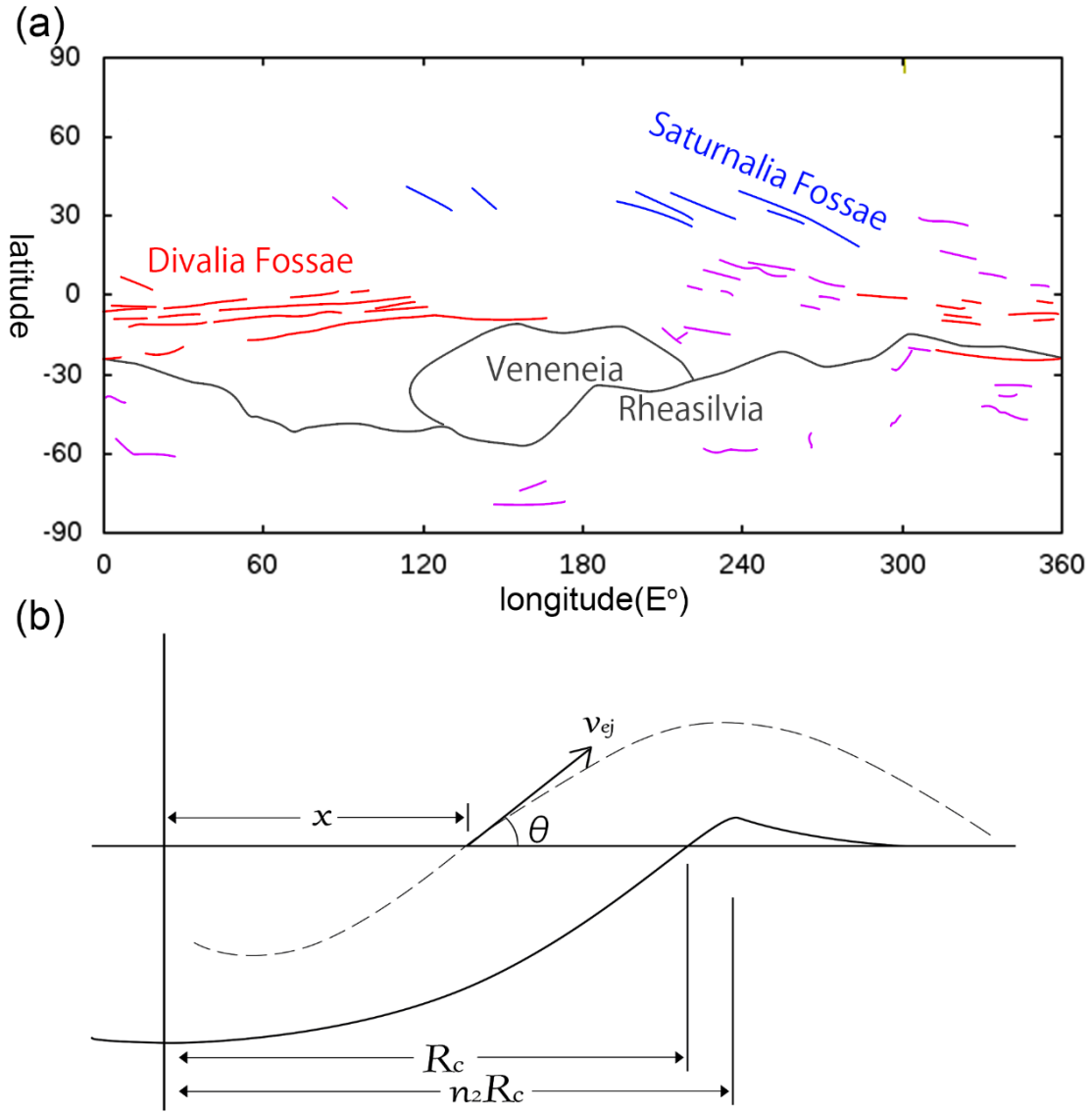

**Figure 1.** (a) Distribution of large-scale troughs, Divalia Fossae (red lines) and Saturnalia Fossae (blue lines) on Vesta and rim crests of Veneneia and Rheasilvia impact basins (black lines). Mapping of troughs and rims are based on a geologic map by Yingst et al. [2014]. The purple lines are structures mapped as troughs by Yingst et al. that are not Divalia or Saturnalia Fossae. (b) Definitions of parameters, adopted from Housen and Holsapple [2011], where $v_{ej}$ is the initial launch velocity of a particle, $R_c$ is the apparent crater radius, $n_2 R_c$ is the crater rim radius, $\theta$ is the launch angle, and $x$ is the launch point (i.e., the distance between the launch position and the impact point).

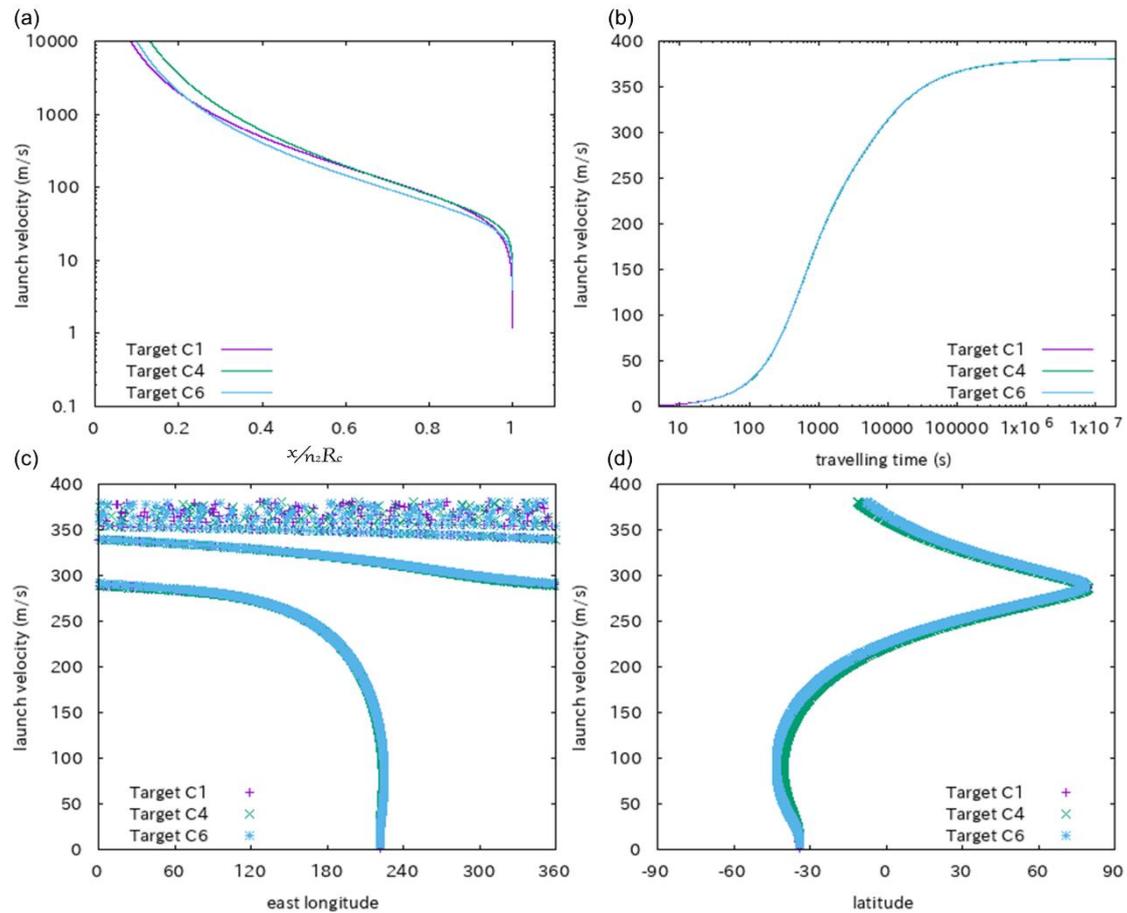

**Figure 2.** (a) Initial launch velocities, $v_{ej}$, as a function of the launch point, $x$, relative to the crater radius, $n_2 R_c$ = 250 km. (b) The relationship between the initial launch velocity and the travelling time (a time from launch to landing) of ejecta particles. (c) The relationship between the initial launch velocity and the longitude of landing locations of ejecta particles indicated with a light blue line drawn in Figs. 3c, 4c, and 5c. (d) The relationship between the initial launch velocity and the latitude of landing locations of ejecta particles indicated with a light blue line drawn in Figs. 3c, 4c, and 5c.

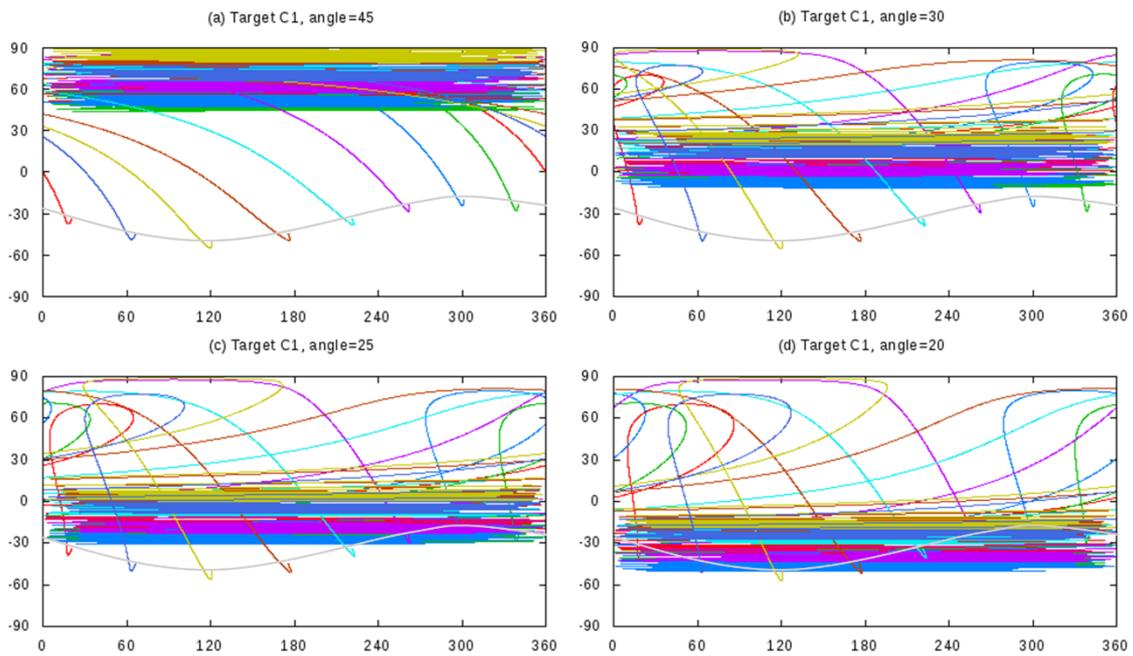

**Figure 3.** The landing locations of particles launched from the Rheasilvia in the case of an initial launch velocity of Target C1 (water). The horizontal and vertical axes are the east longitude and latitude of Vesta. A line with the same color indicates the landing locations of particles ejected in the same direction. Gray lines indicate the rim of Rheasilvia. The initial launch directions are evenly spaced at angles of 45° from the crater center. Initial launch angles were set as (a) $\theta=45°$, (b) $\theta=30°$, (c) $\theta=25°$, and (d) $\theta=20°$.

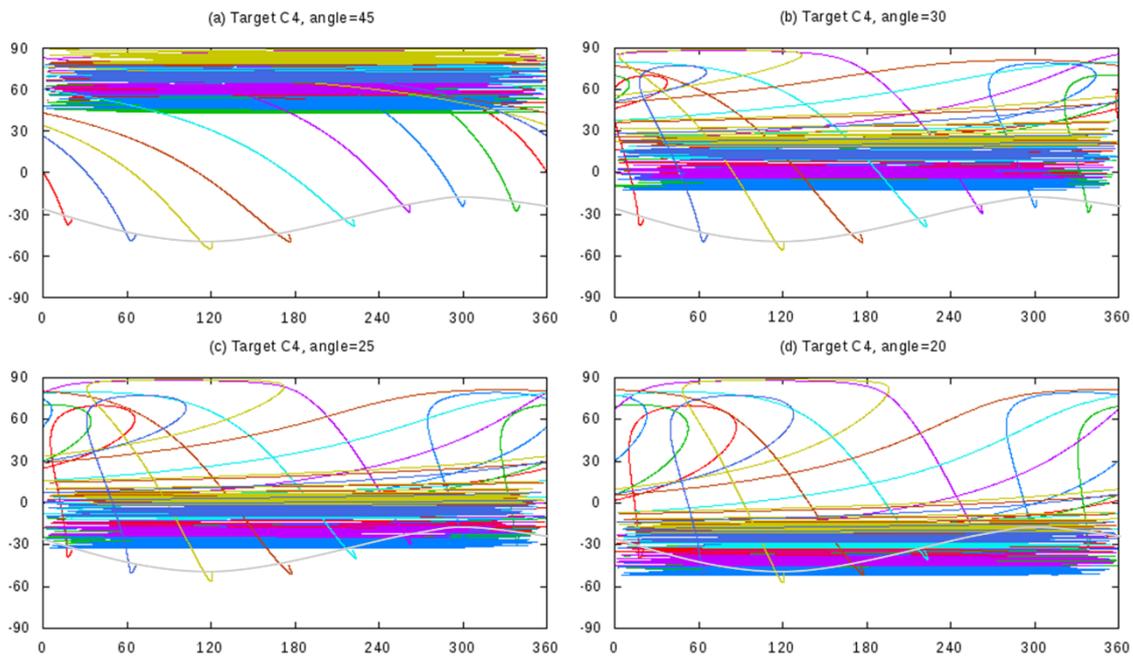

Figure 4. The landing locations of particles launched from the Rheasilvia in the case of an initial launch velocity of Target C4 (dry sand). Initial launch angles were set as (a) $\theta=45°$, (b) $\theta=30°$, (c) $\theta=25°$, and (d) $\theta=20°$.

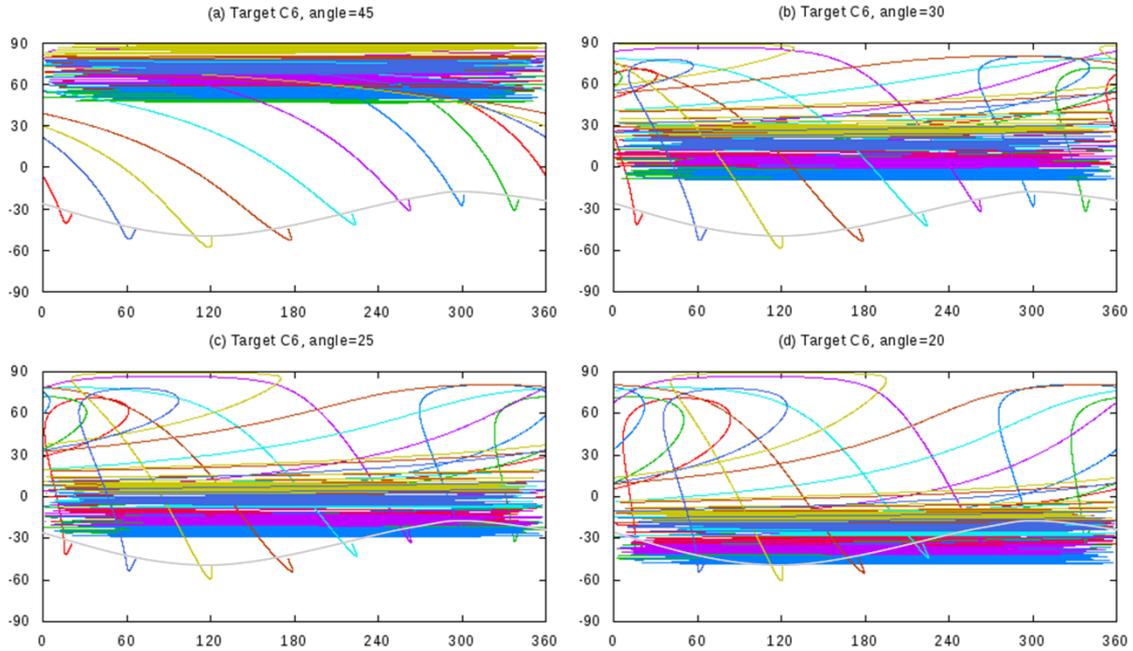

Figure 5. The landing locations of particles launched from the Rheasilvia in the case of an initial launch velocity of Target C6 (glass micro-spheres). Initial launch angles were set as (a) $\theta=45°$, (b) $\theta=30°$, (c) $\theta=25°$, and (d) $\theta=20°$.

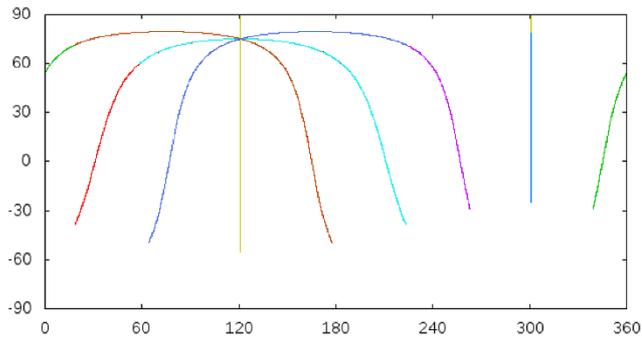

Figure 6. The landing locations of particles launched from the Rheasilvia in the case of an initial launch velocity of Target C4 and initial launch angle of $\theta=45°$. Here, we assumed the asteroid rotation of $T=\infty$ (non-rotation).

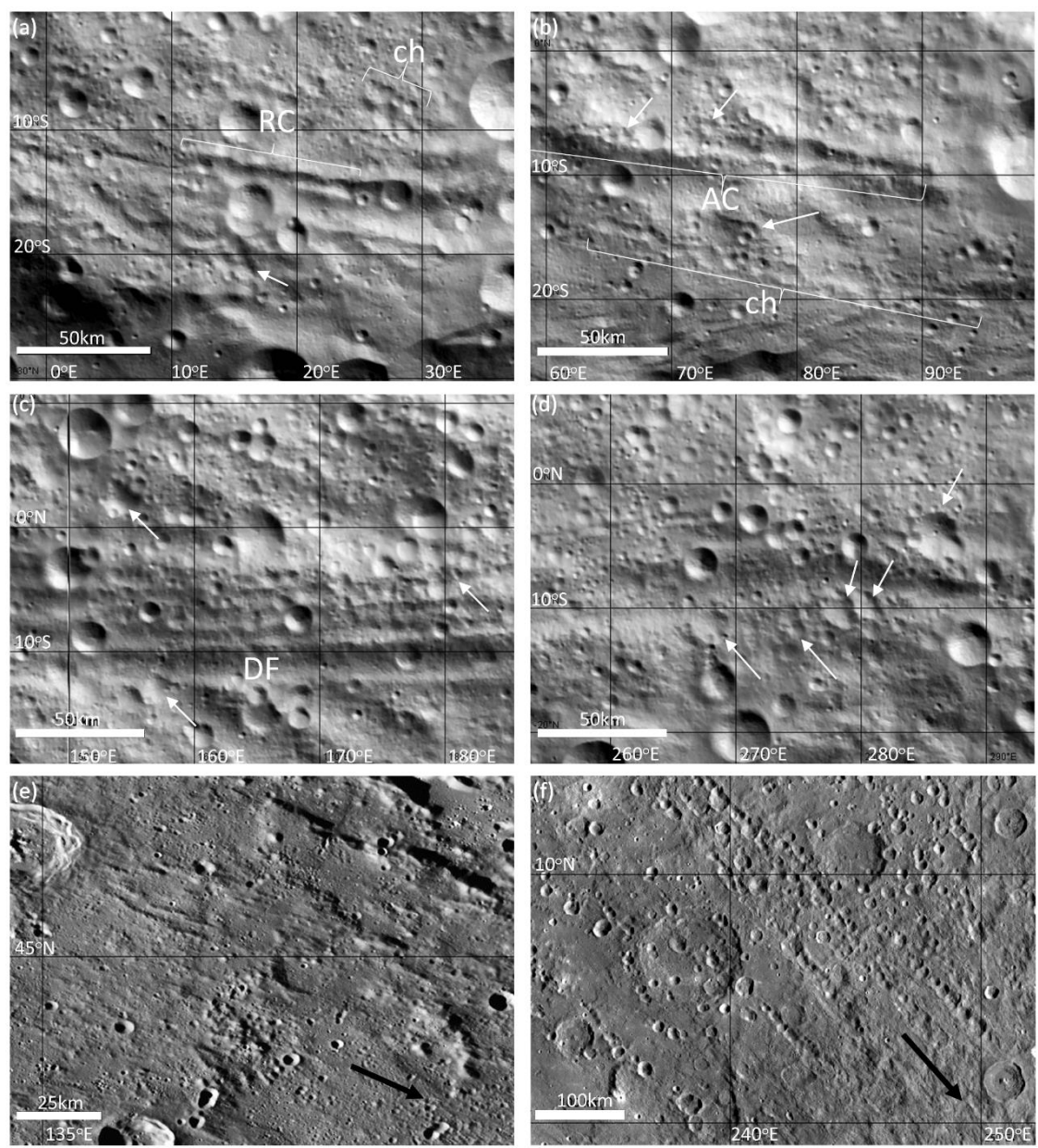

**Figure 7**. Equatorial structures on Vesta. (a) The pit crater chain Robigalia Catena (RC) and an impact crater chain (ch). (b) The pit crater chain Albalonga Catena (AC) and an impact crater chain (ch). Albalonga Catena extends to the west of this image. (c) Example Divalia Fossae troughs, including part of Divalia Fossa (DF). (d) Crater clusters around the equator (white arrows) and a part of Divalia Fossae. (e) A portion of radial sculptures of the Caloris basin. Black arrow shows a direction toward the basin center. (f) A portion of secondary craters of the Orientale basin. The black arrow

shows the direction of the basin center. Base maps are from global mosaics of Vesta (Vesta_Dawn_HAMO_ClrShade_DLR_Global_48ppd.tif), Mercury (Mercury_MESSENGER_MDIS_Basemap_BDR_Mosaic_Global_166m.tif), and the Moon (Lunar_LRO_LROC-WAC_Mosaic_global_100m_June2013.tif) released via the U.S. Geologic Survey (https://planetarymaps.usgs.gov/mosaic/). North is up in all of these images.

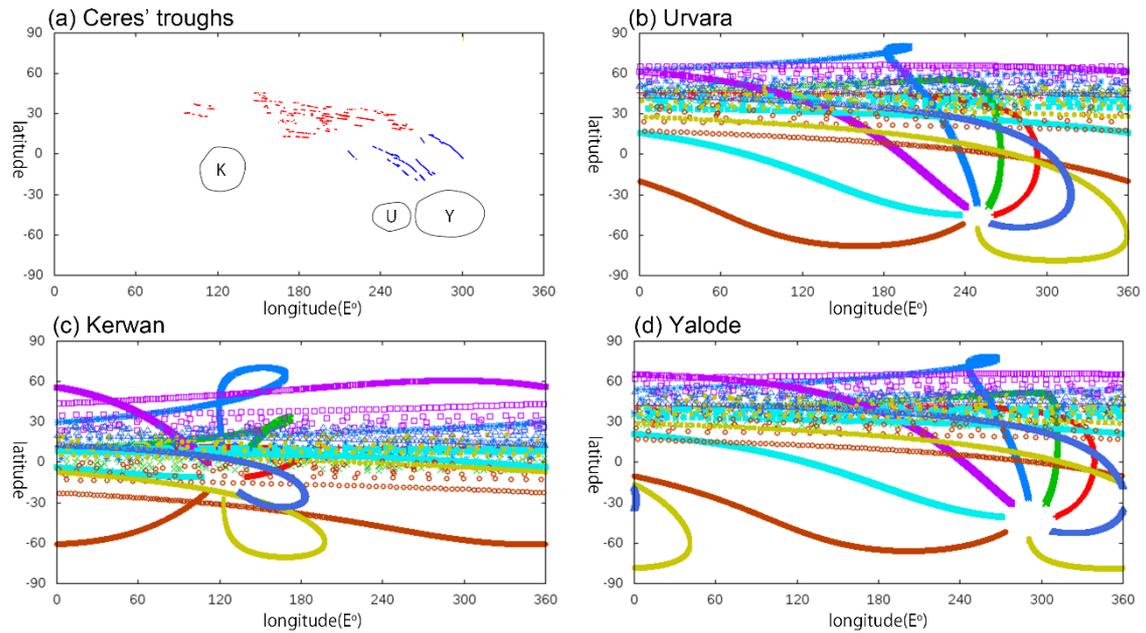

**Figure 8.** (a) The distribution of linear structures on Ceres. Red lines outline the Junina Catenae; blue lines, the Samhain Catenae; and black lines, crater rim crests of Kerwan, Urvara, and Yalode. Mapping of the troughs is derived from a geological map by Scully et al. [2017]. The landing location of ejecta particles from (b) Urvara, (c) Kerwan, and (d) Yalode. We assumed an initial launch velocity of $v_{ej}$ in Eq. (1), scaling constants for C4, and the launch angle of $\theta$=45° in every plate.

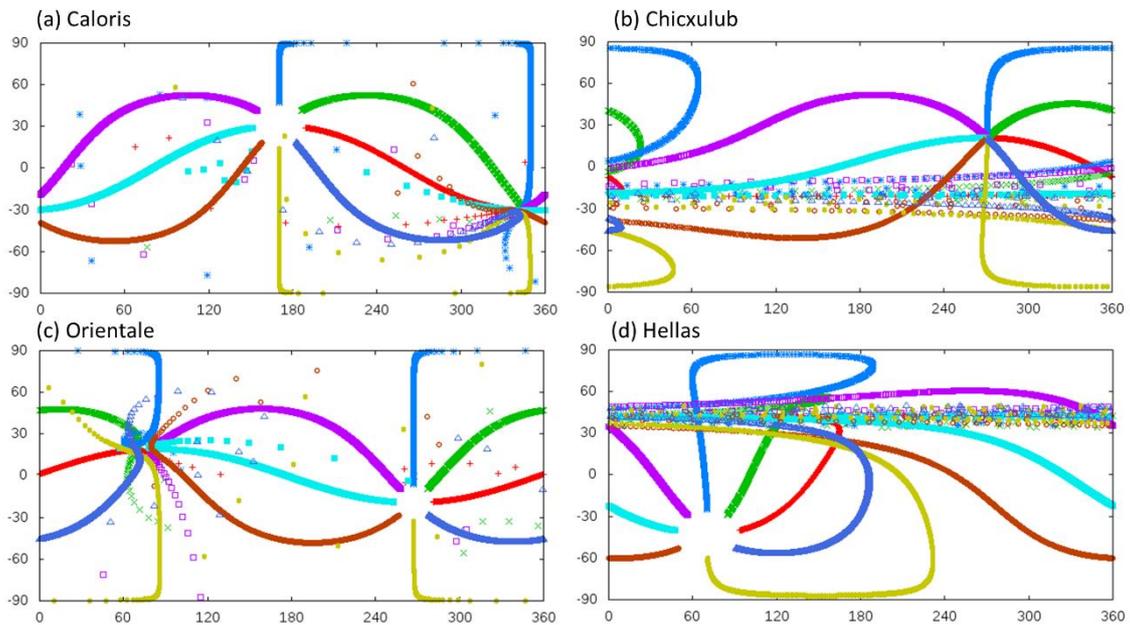

**Figure 9.** The landing locations of ejecta particles from (a) the Caloris on Mercury, (b) the Chicxulub crater on the Earth, (c) the Orientale basin on the Moon, and (d) the Hellas on Mars. In the calculation of (a), the trajectories of ejecta particles are calculated using a combination of the Hill's equation and a rotation reference frame with Mercury's synodic rotation period. In the calculation of (c), the trajectories of ejecta particles are calculated using the Hill's equation. We use a scaling constant of C4 and a launch angle of $\theta=45°$.

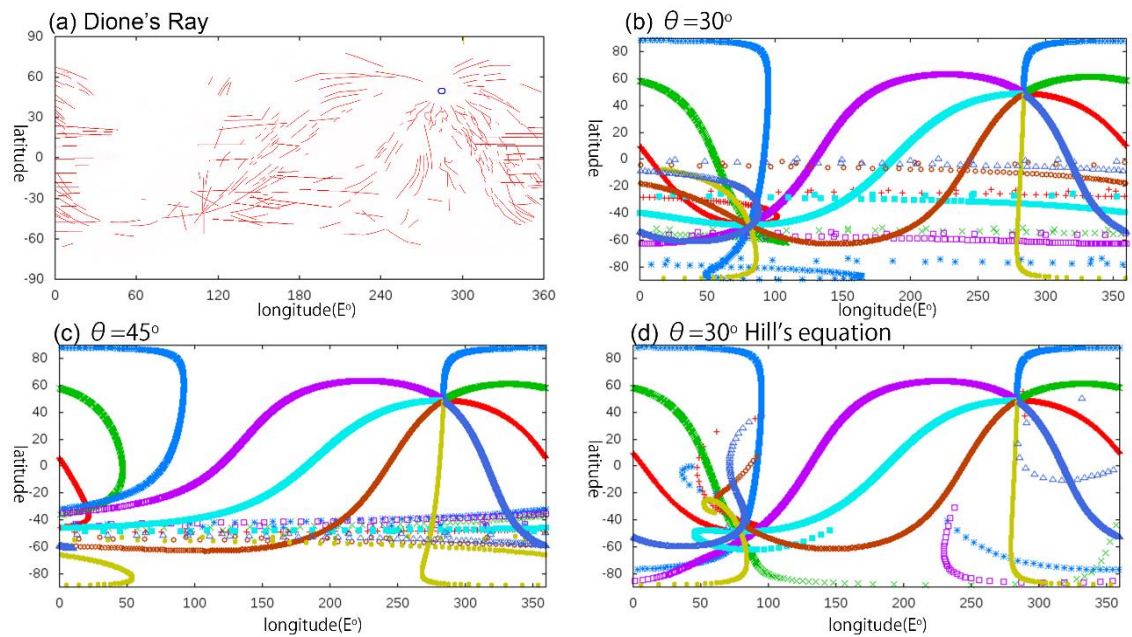

Figure 10. (a) The distribution of rays (red lines) and the Creusa crater (blue circle) on Dione. Mapping of the rays and craters is derived from Hirata and Miyamoto [2016]. Rays from the Creusa crater extend over Dione and the rays on the trailing hemisphere were partially erased by accumulation of the so-called dark materials. Here, we show three calculations for the landing location of ejecta particles from the Creusa crater (b, c, d). The Hill sphere in the calculations of cases of (b) and (c) is enlarged by 6 times the Hill spheres of Dione. In the calculation of (d), the Hill sphere is set as that of the actual Dione and the trajectories of ejecta particles are calculated using the Hill's equation. We use a scaling constant of C4 and a launch angle of (b, d) $\theta=30°$ and (c) $\theta=45°$.

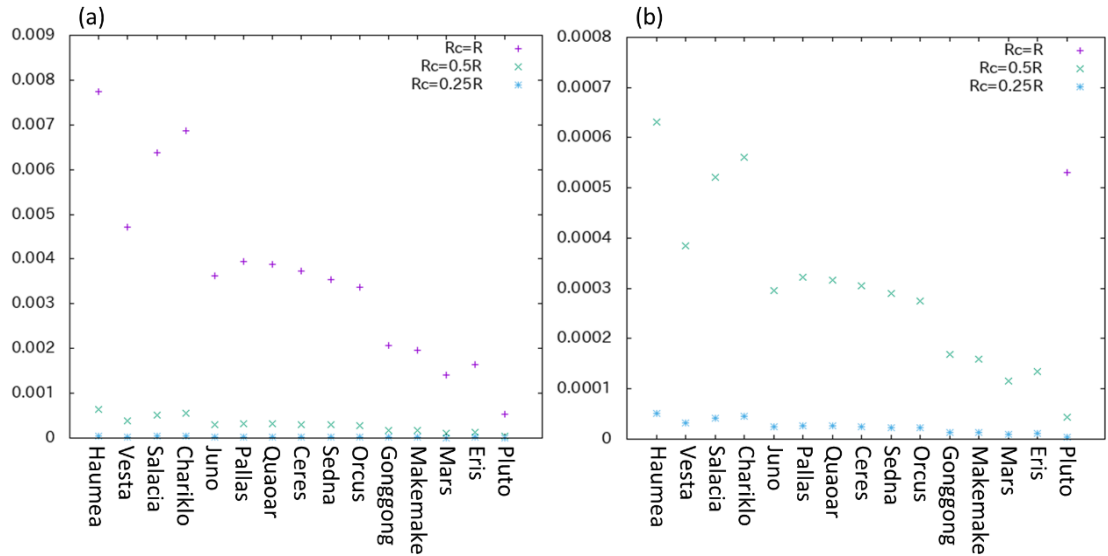

Figure 11. The ratio of the volume of ejecta forming trough system along the latitude to the volume of the object. The right plate (b) is an enlargement of the lower part of the left plate (a). Here, we show a case of a scaling constant of Target C4. The list is sorted by an object with the fastest rotation period (Table 3).

**Table 1. Scaling constants used in the ejecta model.**

|  | C1 | C4 | C6 |
|---|---|---|---|
| $\mu$ | 0.55 | 0.41 | 0.45 |
| $k$ | 0.2 | 0.3 | 0.5 |
| $C_1$ | 1.5 | 0.55 | 1.00 |
| $H_1$ | 0.68 | 0.59 | 0.8 |
| $n_2$ | 1.5 | 1.3 | 1.3 |
| $p$ | 0.5 | 0.3 | 0.3 |

**Table 2 The value of $V_{trough}$ for various impact basins in the solar system[*1].**

| Basin | Object | Basin Dia.[*2] (km) | $v_{max}$ (m/s) | A case of 1 rotation period[*3] | | | A case of 2 rotation period[*4] | | |
|---|---|---|---|---|---|---|---|---|---|
| | | | | $v_{min}$ (m/s) | $V_{trough}$ (km³) | $V_{trough}/V_{object}$ | $v_{min}$ (m/s) | $V_{trough}$ (km³) | $V_{trough}/V_{object}$ |
| **Caloris** | Mercury | 1525 | 4228 | - | 0 | 0 | - | 0 | 0 |
| **Chicxulub** | Earth | 180 | 11167 | 10764 | 1.1x10² | 9.9x10⁻¹¹ | 10922 | 6.4x10¹ | 5.9x10⁻¹¹ |
| **Orientale** | Moon | 930 | 2342 | - | 0 | 0 | - | 0 | 0 |
| **Hellas** | Mars | 2300 | 5022 | 4817 | 1.8x10⁶ | 1.1x10⁻⁵ | 4894 | 1.1x10⁶ | 6.7 x10⁻⁶ |
| **Rheasilvia**[*5] | Vesta | 500 | 363 | 318 | 1.2x10⁵ | 1.5x10⁻³ | 335 | 6.8x10⁴ | 8.9x10⁻⁴ |
| **Veneneia** | Vesta | 400 | 363 | 318 | 5.2x10⁴ | 6.8x10⁻⁴ | 335 | 3.0x10⁴ | 4.0x10⁻⁴ |
| **Kerwan** | Ceres | 284 | 517 | 465 | 8.3x10³ | 1.9x10⁻⁵ | 484 | 4.9x10³ | 1.1x10⁻⁵ |
| **Herschel** | Mimas | 145 | 130 | - | 0 | 0 | - | 0 | 0 |
| **Odysseus** | Tethys | 450 | 341 | - | 0 | 0 | - | 0 | 0 |
| **Turgis** | Iapetus | 580 | 566 | - | 0 | 0 | - | 0 | 0 |
| **Sputnik** | Pluto | 1400 | 1210 | 1191 | 2.1x10⁵ | 3.0x10⁻⁵ | 1198 | 1.3x10⁵ | 1.8x10⁻⁵ |

*1 Here, we present a case of the scaling constant of Target C4.

*2 The basin diameters are from Fassett et al. [2009] (Caloris), Kring [1995] (Chicxulub), Guo et al. [2018] (Orientale), Bernhardt et al. [2016] (Hellas), Schenk et al. [2012] (Rheasilvia and Veneneia), Williams et al. [2018] (Kerwan), Dones et al. [2009] (Herschel and Turgis), Jaumann et al. [2009] (Odysseus), and Schenk et al. [2015] (Sputnik).

*3 In this case, $v_{min}$ is assumed to be the launch velocity of an ejecta particle whose travelling time is comparable to one object rotation period.

*4 In this case, $v_{min}$ is assumed to be the launch velocity of an ejecta particle whose travelling time is comparable to two object rotation period.

*5 The value for Rheasilvia is slightly different from that in Section 4.1, because of the definitions of $v_{min}$ and $v_{max}$.

**Table 3 Physical characteristics of various solar system objects *1.**

|  |  | Rotation Period (h) | Mass ($10^{21}$kg) | Diameter (km) | a (AU) | $v_{min}$*2 (m/s) | $v_{max}$ (m/s) |
|---|---|---|---|---|---|---|---|
| 136108 | Haumea | 3.9 | 4.01 | 1590 | 43.1 | 668 | 821 |
| 4 | Vesta | 5.34 | 0.259 | 525 | 2.4 | 318 | 363 |
| 120347 | Salacia | 6.1 | 0.492 | 914 | 42.3 | 219 | 379 |
| 10199 | Chariklo | 7 | 0.007 | 258 | 15.8 | 71 | 85 |
| 3 | Juno | 7.21 | 0.0286 | 246 | 2.7 | 159 | 176 |
| 2 | Pallas | 7.81 | 0.214 | 545 | 2.8 | 289 | 324 |
| 50000 | Quaoar | 8.8 | 1.4 | 1083 | 43.6 | 526 | 588 |
| 1 | Ceres | 9.07 | 0.94 | 939 | 2.8 | 465 | 517 |
| 90377 | Sedna | 10.27 | 1 | 995 | 499.5 | 468 | 518 |
| 90482 | Orcus | 13.2 | 0.634 | 965 | 39.1 | 380 | 419 |
| 225088 | Gonggong | 22.4 | 1.75 | 1230 | 67.5 | 580 | 617 |
| 136472 | Makemake | 22.8 | 3.1 | 1430 | 45.3 | 71 | 761 |
|  | Mars | 24.62 | 641.71 | 6779 | 1.5 | 4817 | 5022 |
| 136199 | Eris | 25.9 | 16.46 | 2326 | 68.0 | 1310 | 1375 |
| 134340 | Pluto | 153.3 | 13.03 | 2377 | 39.4 | 1191 | 1210 |

*1 Here, we present a case of the scaling constant of Target C4. The mass of Juno is from Baer et al. [2011], mass and diameter of Chariklo from Leiva et al. [2017], mass of Orcus and Salacia from Grundy et al. [2019], diameter of Orcus, Salacia, and Quaoar from Brown and Butler [2017], mass of Haumea from Ragozzine and Brown [2009], and diameter of Haumea from Ortiz et al. [2017], mass of Quaoar from Fraser et al. [2013], mass and diameter of Gonggong from Kiss et al. [2019], mass of Eris from Holler et al. [2021], diameter of Eris from Sicardy et al. [2011], mass of Makemake from Parker et al. [2018], diameter of Makemake from Brown [2013], diameter of Sedna from Pál et al. [2012], diameter and mass of Pluto from Stern et al. [2015], mass of Sedna is obtained by assuming its density to be 2000 kg/m³, and the rest from online archive via Jet Propulsion Laboratory Small-Body Database Browser.

*2 In this case, $v_{min}$ was assumed to be the launch velocity of ejecta particle whose travelling time is comparable to one object rotation period.

**Appendix A. Cases where the launch angle is not constant.**

In the main text, all calculations assume that the launch angle, $\theta$, is constant with respect to the launch position, $x$. Here, cases were evaluated where the launch angle is variable with respect to x. As shown in the left plates in Fig. A1, we generated several examples of $\theta$ as a function of $x$ using a random generator. Right plates in Fig. A1 are the results of the landing locations of ejecta particles assuming a function of $\theta$ on their left sides. As a result, it appears that the landing locations exhibit slightly sinuous shapes, but their pattern is approximately along the latitude line. It is clear that if the launch angle is varied more significantly, a chaotic pattern is exhibited; however, as long as the launch angle is somewhat continuous, a pattern along the latitude line can be obtained. We assume a scaling constant of C4.

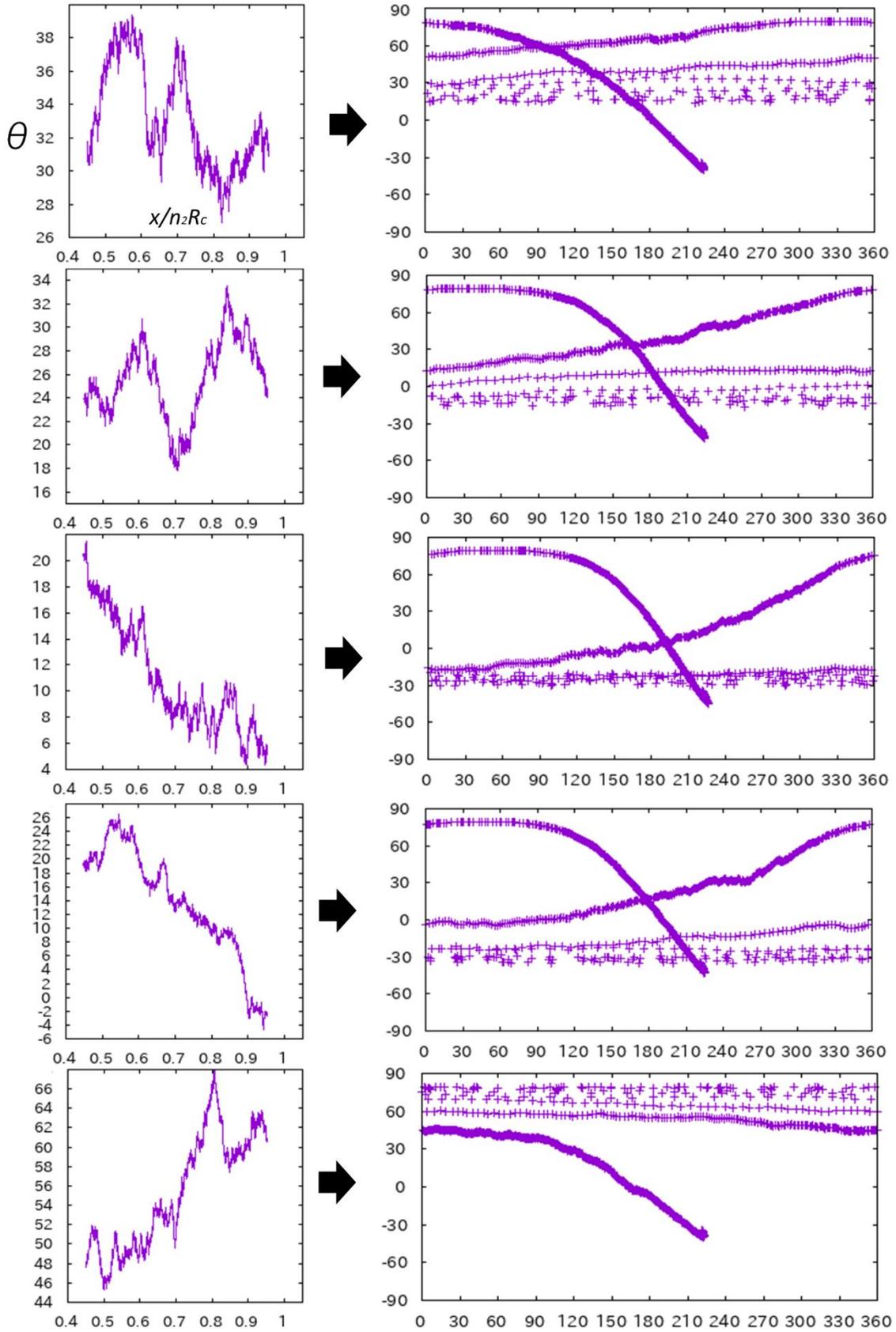

**Figure A1.** Left plates show examples of $\theta$ (vertical) as a function of $x$ (horizontal), and right plates show the latitude (vertical) and east longitude (horizontal) of landing locations of ejecta particles assuming a function of $\theta$ in their left plates.

## Appendix B. Origin of a trough system that encircles roughly 2/3 of Vesta's equatorial region

Our model does not suggest the formation of trough systems that completely encircle the equator of Vesta. Landing locations of ejecta particles incompletely covering the equator can occur, assuming that the initial launch velocity distribution and/or launch azimuth direction are asymmetric. As an example, Fig. B1 shows the landing locations of particles launched from the Rheasilvia in the case of initial launch velocities between 326m/s and 335m/s, launch azimuth angles between 225° and 360°, and an initial launch angle of 25°. In this case, the ejecta particles did not accumulate on Vestalia Terra. Various initial launch conditions can create a similar distribution. When the initial launch velocity or launch azimuth of the ejecta particles are inhomogenous, it is possible to create various distributions with intermittent or incomplete chains. It is also possible to create a distribution in which ejecta particles accumulate only on Vestalia Terra. Various incomplete distributions can be created by providing a certain range of launch velocity distributions and/or asymmetric azimuthal directions. This assumption is not unnatural. In fact, secondary craters or chains distributed far away from the impact basin, such as Imbrium or Orientale, are asymmetric and concentrated in a certain direction or at a certain distance, which indicates that the initial launch velocity or azimuth of the secondary impactors are not homogenous.

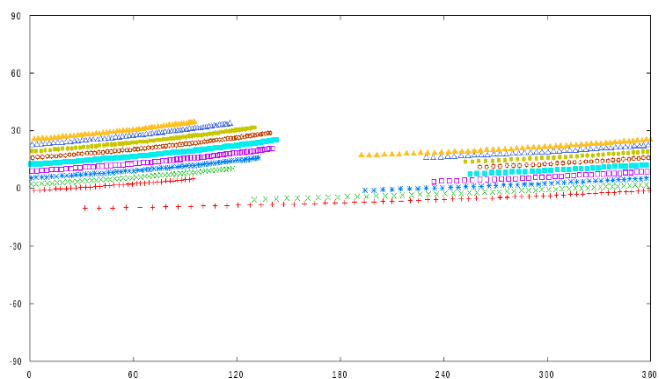

**Figure B1.** The landing locations of particles launched from the Rheasilvia in the case of an initial launch velocity of Target C4, initial launch velocities between 326m/s and 335m/s, launch azimuth angles between 225° and 360°, and an initial launch angle of 25°.